\pdfoutput=1

\documentclass[11pt,aps,onecolumn,preprint,nofootinbib]{revtex4}

\usepackage{times}

\usepackage{amsmath}
\usepackage{graphicx}
\renewcommand{\L}{\mathcal{L}}

\newcommand{\figref}[1]{FIG.~\ref{#1}}

\newcommand{\vecx}{\mathbf{x}}

\newcommand{\best}{\mathrm{best}}

\usepackage{bibunits}

\begin{document}

\title{Efficient inference of parsimonious phenomenological models of
  cellular dynamics using S-systems and alternating regression}


\author{Bryan C.~Daniels$^{\ast}$}
\affiliation{Center for Complexity and Collective Computation, Wisconsin Institute for Discovery,
University of Wisconsin, Madison, WI 53715, USA}

\author{Ilya Nemenman$^{\ast}$}
\affiliation{Departments of Physics and Biology\\
Emory University, Atlanta, GA 30322, USA\\}

\begin{abstract}
  The nonlinearity of dynamics in systems biology makes it hard to
  infer them from experimental data.  Simple linear models are
  computationally efficient, but cannot incorporate these important
  nonlinearities.  An adaptive method based on the S-system formalism,
  which is a sensible representation of nonlinear mass-action kinetics
  typically found in cellular dynamics, maintains the efficiency of
  linear regression.  We combine this approach with adaptive model
  selection to obtain efficient and parsimonious representations of
  cellular dynamics.  The approach is tested by inferring the dynamics
  of yeast glycolysis from simulated data. With little computing time,
  it produces dynamical models with high predictive power and with 
  structural complexity adapted to the difficulty of the inference
  problem.
  
  $^\ast$E-mail: bdaniels@discovery.wisc.edu,
    ilya.nemenman@emory.edu

\end{abstract}

\maketitle



\section{Introduction}
  
Dynamics of cellular regulation are driven by large and intricate
networks of interactions at the molecular scale. Recent years have
seen an explosive growth in attempts to automatically infer such
dynamics, including their functional form as well as specific
parameters, from time series of gene expression, metabolite
concentrations, or protein signaling data
\cite{Wiggins:2003ut,Chou:2006dh,Schmidt:2009dt,Lillacci:2010jm,Schmidt:2011gf}.
Searching the combinatorially large space of all possible multivariate
dynamical systems requires vast amounts of data (and computational
time), and realistic experiments cannot sufficiently constrain many
properties of the inferred dynamics \cite{Gutenkunst:2007gl}.  Thus
detailed, mechanistic models of cellular processes can overfit and are
not predictive. Instead, one often seeks {\em a priori} information to
constrain the search space to simpler {\em ad hoc} models of the
underlying complex processes
\cite{Chou:2006dh,Lillacci:2010jm,Neuert:2013ed}. This can lead to
missing defining features of the underlying dynamics, and hence to
poor predictions as well.

A promising approach is to forgo mechanistic accuracy in favor of {\em
  phenomenological} or {\em effective} models of the dynamics
\cite{Wiggins:2003ut,Daniels:2014vz}. Following the recipe that can be
traced to Crutchfield and McNamara \cite{Crutchfield:1987va}, one
defines a sufficiently general (and, hopefully, complete
\cite{Nemenman:2005cw}) set of possible dynamics using pre-defined
combinations of basis functions. One then uses statistical model
selection techniques to select a model that is ``as simple as
possible, but not simpler'' than needed for predictive modeling of the
studied system \cite{Hastie:2009wp}. Such approaches expand
traditional statistical modeling to the realm of nonlinear dynamical
systems inference. While their derived models may not have an easy
mechanistic interpretation, they are parsimonious and often predict
responses to yet-unseen perturbations better than microscopically
accurate, and yet underconstrained models.

In particular, a recent successful approach of this type
\cite{Daniels:2014vz} used S-systems (to be described later in this
article) as a basis for representing the dynamics. The
representation is nested, so that the dynamics can be ordered by an
increasing complexity. It is also complete, so that every sufficiently
smooth dynamics can be represented with an arbitrary accuracy using
large dynamical S-systems, which grow to include complex
nonlinearities and dynamical variables unobserved in experiments. Thus
inference of the dynamics with this representation is statistically
consistent \cite{Nemenman:2005cw} in the sense that, for a large
amount of data, the inferred models are guaranteed to become
arbitrarily accurate with high probability. The success of such
phenomenological modeling is evident from its ability to use simulated
noisy data to infer that equations of motion in celestial mechanics
are second order differential equations, and to derive Newton's law of
universal gravitation in the process \cite{Daniels:2014vz}. Similarly,
the approach infers parsimonious, phenomenological, and yet highly
accurate models of various cellular processes with many hidden
variables using about $500$ times fewer data than alternative methods
that strive for the mechanistic accuracy \cite{Daniels:2014vz}.

And yet the approach in \cite{Daniels:2014vz} suffers from a high
computational complexity, requiring solutions of millions of trial
systems of differential equations to generalize well. In specific
scenarios, it may be useful to trade some of its generality for a
substantial computational speedup. As has been argued in
\cite{Chou:2006dh,Schmidt:2011gf}, in biological experiments, it is
often possible to measure not just expressions of all relevant
molecular species, but also their rates of change. In this article, we
show that, in such a scenario, the alternating regression method for
inference of S-systems \cite{Chou:2006dh}, merged with the Bayesian
model selection for dynamics \cite{Daniels:2014vz}, infers dynamical
models with little computational effort. Importantly, these
phenomenological models are parsimonious, nonlinear, and hence
predictive even when the microscopically accurate structure of the
dynamics is unknown.

In the remainder of the article, we first formalize the problem we
study, then introduce the reader to S-systems formalism. We then
explain the alternating regression method for inference of the
S-systems parameters from data, followed by our implementation of
S-systems model selection using Bayesian approaches. Finally, we
introduce the biological system on which our approach is tested
\cite{RuoChrWol03}, and present the results of the automated dynamical
systems inference in this context.

\section{Problem setup: Inferring biochemical dynamics from data}
In a general scenario \cite{Daniels:2014vz}, one needs to infer the
deterministic dynamics from a time series of $J$-dimensional vectors
of molecular expressions, $\{x_\mu\}_i=\vecx_i$, $\mu=1\dots J$,
measured with noise at different time points $t_i$, $i=1\dots N$. The
system may be arbitrarily nonlinear and the measured data vectors
$\vecx$ may represent only a small fraction of the total number of
densely interacting dynamical variables. Intrinsic noise in the
system, while possibly important \cite{Raj:2008ip}, is neglected for
simplicity.

We focus on a simplified case \cite{Chou:2006dh,Schmidt:2011gf} in
which the measured dynamical variables completely define the system,
and where their rates of change, $d\vecx_i/dt=\vecx'_i$, and the
expected experimental variances of the rates,
${\boldsymbol\sigma'_i}^2$, are also given. Since typical experiments
measure variables with higher accuracy than their rates of change, for
simplicity, we assume no errors in $\vecx_i$.\footnote{This framework
  can also be straightforwardly extended to include known exogenous
  variables; for simplicity we do not include this in the example
  here.}  Since the mechanistically accurate form of ${\mathbf X}'$ is
unknown, we are required then to {\em approximate} the rate function
${\mathbf X}'$:
\begin{eqnarray}
  \frac{d\vecx}{dt} &=& {\mathbf X}'(\vecx).\label{eq:dynamics}
\end{eqnarray}
Knowing $\vecx'_i$,
we can fit the functions ${\mathbf X}'$ by minimizing
\begin{equation}
\label{derivChiSq}
\chi^2 =
    \sum_{i=1}^N\sum_{\mu=1}^J
        \left( \frac{x'_{i,\mu} - X'_\mu(\vecx_i)}{\sigma'_{i,\mu}} \right)^2,
\end{equation}
subject to a traditional penalty for the model complexity.  Note that
here we use the measured values $\vecx_i$ instead of the integrated
model values $\tilde \vecx(t_i)$ as the argument of ${\mathbf X}'$.

Different approximation strategies correspond to different functional
forms of ${\mathbf X}'$. In this paper, we focus on the linear
approximation,
\begin{equation}
  \frac{dx_\mu}{dt}=\sum_{\nu=1}^J A_{\mu\nu}x_\nu,\label{eq:linear}
\end{equation}
as well as on S-systems (see below), which can be viewed as linear
models in logarithmic space. However, compared to the previous
work \cite{Chou:2006dh}, the analyzed dynamics is not expected to be
fitted exactly by either of the model families, resulting in
approximate, effective, or phenomenological models. Further, even for
a modestly large number of variables, the number of interaction
parameters, such as $A_{\mu\nu}$, can be quite large. Thus we
additionally study the effects of constraining the interaction model
using Bayesian model selection.

Our {\bf goals} are to understand (1) the accuracy afforded by the
nonlinear S-system approach for phenomenological modeling, especially
compared to a simpler linear regression, (2) the computational
efficiency of the S-systems alternating regression method compared to
more complex fitting approaches in
\cite{Schmidt:2011gf,Daniels:2014vz}, and (3) whether selection of
parsimonious models using the Bayesian approach, first tried for
dynamical systems in Ref.~\cite{Daniels:2014vz}, similarly provides an
improvement over fitting a complex, fully connected interaction model.

\section{S-systems formalism}

The textbook representation of biochemical dynamics uses ordinary
differential equations in the {\em mass-action} form, where the rate
laws are differences of production ($G$) and degradation ($H$) terms,
\begin{equation}
\label{generalPowerlaw}
\frac{dx_\mu}{dt} = G_\mu(\vecx)- H_\mu(\vecx).
\end{equation}
In their turn, $G$ and $H$ are products of integer powers of
expressions of other chemical species, where the powers represent
stoichiometry. For example, if a certain chemical $\mu$ is produced by
a bimolecular reaction involving one molecule of $\nu$ and two
molecules of $\lambda$, then its mass-action production term is
$G_\mu=\alpha_\mu x_\nu x_\lambda^2$, where $\alpha_\mu$ is the
reaction rate.

In what became known as the S-systems formalism, Savageau and
colleagues generalized this form to non-integer powers of expressions
\cite{Savageau:1987vt}. That is,
\begin{eqnarray}
\label{powerLawTerms}
G_\mu(\vecx) = \alpha_\mu \prod_{\nu=1}^{J} x_\nu^{g_{\mu\nu}}, \;
H_\mu(\vecx) = \beta_\mu  \prod_{\nu=1}^{J} x_\nu^{h_{\mu\nu}}.
\end{eqnarray}
The S-system is a {\em canonical representation} of nonlinear
biochemical dynamics since, in a process called {\em recasting}, any
dynamics, Eq.~(\ref{eq:dynamics}), with ${\mathbf X}'$ written in
terms of elementary functions can be rewritten in this power-law form
by introducing auxiliary dynamical variables and initial value
constraints in a certain way \cite{Savageau:1987vt}.  Further, since
any sufficiently smooth function can be represented as a series of
elementary functions (e.\ g., Taylor series), a recasting into an
S-system of a sufficient size can approximate any such deterministic
dynamics. While a detailed review of recasting is beyond the scope of
this article, here we give a few examples. First, the classic
Michaelis-Menten enzymatic kinetics, $x'_1=Ax_1/(B+x_1)$, can be
recast by introducing one hidden variable $x_2$ as
\begin{align}
x'_1&=Ax_1x_2^{-1},\;x_2=B+x_1.
\end{align}
Similarly, the dynamics $x'_1=\sin x_1$ has a representation
\begin{align}
x'_1=x_2,\;x'_2=x_3x_2,\;x'_3=-x_2^2,
\end{align}
where $x_2=\sin x_1$, and $x_3=\cos x_1$. Note that, since the
exponents are not constrained to be positive or integer, dynamics in
this class are generally ill-defined when variables are not positive.

The power-law structure of $G$ and $H$ takes a particularly simple form
if viewed in the space of logarithms of the dynamical variables,
$\xi_\mu=\log x_\mu$:
\begin{eqnarray}
\label{logPowerLawTerms}
\log G_\mu(\vecx) = \log \alpha_\mu + \sum_{\nu=1}^{J} {g_{\mu\nu}}\xi_\nu, \;\;
\log H_\mu(\vecx) = \log \beta_\mu  + \sum_{\nu=1}^{J} {h_{\mu\nu}}\xi_\nu.\label{eq:loglinear}
\end{eqnarray}
Thus S-systems can be rationalized even when the set of the modeled
variables is fixed, not allowed to be enlarged, and hence recasting is
impossible: S-systems are the first order expansion of the logarithms
of the production/degradation terms in the logarithms of the
concentration variables. Arbitrary dynamics can be approximated again
by adding a sufficient number of higher order terms in this
logarithmic Taylor expansion, just like they can be added to
Eq.~(\ref{eq:linear}). However, it has been argued that typical
biochemical kinetics laws remain linear for broader ranges in the
logarithmic, rather than the linear space \cite{Savageau:2002th}. At
the same time, they can produce richer dynamics than that of simple
linear models. Thus S-systems have a potential to be more effective in
approximating biological dynamics. Verifying this assertion is one of
the goals of this article.

\section{Alternating regression for S-systems inference}
\label{linearRegression}

The key observation that leads to efficient inference of S-systems is
that each of the production and degradation terms is linear in its
parameters in log-space, \eqref{logPowerLawTerms} \cite{Chou:2006dh}.
Specifically, if we know the current concentrations $\vecx$ and their
time derivatives $\vecx'$, and hold constant the parameters in one of
the two terms (say, $G$), the parameters in the other term ($H$) can
be fit using linear regression.  The alternating regression approach,
first implemented in \cite{Chou:2006dh} for models that can be fitted
exactly by S-systems, simply iteratively switches between
linearly fitting parameters in the two terms.  Thus, for each S-system
model, defined by which interactions $g_{\mu\nu}$ and $h_{\mu\nu}$ are
allowed to be nonzero, the inference consists of the following two
steps repeated until convergence:
\begin{enumerate}
\item \emph{Fit production terms, holding degradation fixed.}  That
  is, in \eqref{generalPowerlaw}, solve for $\alpha_\mu$ and
  $g_{\mu\nu}$ using a linear regression (see section
  \ref{linearRegression}), while holding $\beta_\mu$ and $h_{\mu\nu}$
  fixed.\footnote{Due to the nonlinear nature of the transformation to
    log-space, we are not guaranteed to have a smaller $\chi^2$ in 
    linear space after the linear regression in log-space.
    In practice, we find reliable convergence given a modest
    quantity of data.}
\item \emph{Fit degradation terms, holding production fixed.}  Same,
  but swapping $(\alpha_\mu,g_{\mu\nu})$ and $(\beta_\mu,h_{\mu\nu})$.
\end{enumerate}

To implement this logarithmic linear regression, we define
\begin{equation}
\label{Yeqn}
Y^{(G)} = \log(H + x'), \; Y^{(H)} = \log(G - x'),
\end{equation}
so that in the regression in the two cases we are attempting, we are
looking for parameters $\alpha, \beta, g,$ and $h$ that satisfy, for
every measured timepoint $t_i$,
\begin{eqnarray}
Y^{(G)}_{i,\mu} = \log \alpha_\mu + \sum_\nu g_{\mu\nu} \xi_{i,\nu}, \;
Y^{(H)}_{i,\mu} = \log\beta_\mu + \sum_\nu h_{\mu\nu} \xi_{i,\nu}.
\end{eqnarray}
We define parameter matrices $P$ with a row for the prefactors
$\log\alpha$ and $\log\beta$ followed by the matrix of the exponent
parameters:
\begin{align}
&P^{(G)}_{\kappa=1,\mu} = \log \alpha_\mu, &P^{(G)}_{\kappa=1+\nu,\mu} = g_{\mu\nu}; \\
&P^{(H)}_{\kappa=1,\mu} = \log \beta_\mu, & P^{(H)}_{\kappa=1+\nu,\mu} = h_{\mu\nu}.
\end{align}
Then, for both $H$ and $G$, the problem becomes a linear regression in
which we want to minimize the following modified $\chi^2$ with respect
to $P$:
\begin{equation}
\label{chiyEqn}
\tilde \chi_Y^2 =
    \sum_{\mu, i} \left[ W_{i,\mu} (Y_{i,\mu} 
        - \sum_\kappa D_{i,\kappa} P_{\kappa,\mu}) \right]^2
        + \sum_{\mu,\kappa} \frac{P_{\kappa,\mu}^2}{\varsigma_p^2},
\end{equation}
where the last term corresponds to Gaussian priors on each parameter
with mean $0$ and variance $\varsigma_p^2$ (in our example below,
we set $\varsigma_p = 10^{-1}$).  Here the design matrix
$D$ combines the prefactor parameters and the data,
\begin{equation}
  D_{i,\kappa=1} = 1, \; D_{i,\kappa=1+\nu} = \xi_{i,\nu},
\end{equation}
and the matrix $W$ weights each residual according to its
uncertainty in log-space,
\begin{equation}
\label{weightEqn}
W_{i,\mu} = \frac{1}{\sigma'_{i,\mu}} \frac{d \exp Y_{i,\mu}}{d Y_{i,\mu}}
       = \frac{\exp{Y_{i,\mu}}}{\sigma'_{i,\mu}}.
\end{equation}

Finally, elements of $Y$ are undefined when the arguments of the logs
in \eqref{Yeqn} become negative, corresponding to derivatives that
cannot be fit by modifying only $G$ or $H$ at one time.  In these
cases, we zero the corresponding weight $W_{i,\mu}$, effectively
removing these datapoints from the regression at a particular
iteration. This can be a bigger problem, in principle, for approximate
approaches than it was for the case when S-systems could fit the data
exactly \cite{Chou:2006dh}.  In practice, the number of such
datapoints quickly becomes small or zero after just a few iterations.

The linear regression can be solved separately for each species $\mu$.
In matrix form, extremizing the $\mu$th term of $\tilde \chi_Y^2$ in
\eqref{chiyEqn} produces the maximum likelihood parameters for species
$\mu$:
\begin{equation}
\label{regression}
P_\mu = \left(\check D_\mu^T \check D_\mu 
  + \frac{1}{\varsigma_p^2} I\right)^{-1}
\check D_\mu^T \check Y_\mu,
\end{equation}
where $I$ is the identity matrix, and $(\check D_\mu)_{i,\nu} =
W_{i,\mu} D_{i,\nu}$, $(\check Y_\mu)_{i} = W_{i,\mu} Y_{i,\mu}$.  

To perform the regression with some parameters fixed, it is convenient
to let all matrices remain the same shape instead of removing rows and
columns corresponding to parameters fixed at the current iteration.
To accomplish this, we define the binary matrix $\theta$ that is the
same shape as $P$ and contains a 1 when the corresponding parameter is
to be optimized and a 0 when it is not.  Because in our model the
default fixed value for each parameter is 0, we arrive at $(\check
D_\mu)_{i\nu} = W_{i,\mu} D_{i,\nu} \theta_{\mu,\nu}$ and
\begin{equation}
P_\mu = \left(\check D_\mu^T \check D_\mu
       + \frac{1}{\varsigma_p^2} I
       + (1-\theta_\mu)\delta_{\mu,\nu}\right)^{-1}
      \check D_\mu^T \check Y_\mu,
\end{equation}
where the $\delta_{\mu,\nu}$ term removes singularities corresponding
to the fixed parameters.

In our experiments, the alternating regression typically converged to
a relative tolerance of $10^{-2}$ in a few tens of iterations.  This
made it not much slower than the simple linear regression (and many
orders of magnitude faster than approaches of
Refs.~\cite{Schmidt:2011gf,Daniels:2014vz}), while preserving the
ability to fit nonlinearities pervasive in biological systems.

\section{Adaptive Bayesian dynamical model selection}

Adaptive Bayesian model selection defines an {\em a priori} hierarchy
of models with an increasing complexity and selects the one with the
largest Bayesian likelihood given the data. This minimizes the
generalization error
\cite{MacKay:1992ul,Balasubramanian:1997vr,Nemenman:2005cw}.
Generally, one needs to define such a hierarchy over the number of
interactions among species, the nonlinearity of the interactions, and
the number of unobserved species in the model \cite{Daniels:2014vz}.
Here we have a fixed number of species, and the situation is simpler.

Indeed, different S-systems models can be characterized by ``who
interacts with whom'' --- the identity of those exponential parameters
$g_{\mu\nu}$ and $h_{\mu\nu}$ that are allowed to be nonzero. Then the
complexity hierarchy over S-systems can be defined by gradually
allowing more interactions, so that the production and degradation of
species depends on more and more of (other) species.  Specifically, we
start with the simplest model with 2$J$ parameters in which only
$\beta_\mu$ and $g_{\mu\mu}$ need to be inferred from data. Next
$h_{\mu\mu}$ are added to the list of inferable variables, and then
$\alpha_\mu$ for each $\mu$.  Next are $g_{\mu\mu}$ connections
starting with the nearest neighbors [$\mu-\nu=1 \pmod{J}$], then the
next-nearest neighbors [$\mu-\nu=2 \pmod{J}$], and so on until all
$g_{\mu\nu}$ are included. Finally we add $h_{\mu\nu}$ connections in
the same order.  The final, largest model contains $2J(J+1)$
parameters. While in our examples the order of the variables and hence
the hierarchy is pre-defined, a random order is also possible. As has
been argued elsewhere \cite{Daniels:2014vz}, a random hierarchy is
still better than no hierarchy at all. 


With the hierarchy defined, we calculate the posterior log-likelihood
$\L$ of each model $M$ within the S-systems family in the usual way
\cite{MacKay:1992ul}. We expand the log-likelihood of the model to the
lowest order in $1/N$, that is, as a quadratic form near the maximum
likelihood values determined by the alternating regression in Section
\ref{linearRegression}. This extends the Bayesian Information
Criterion \cite{Sch78} by explicitly incorporating parameter
sensitivities and priors over the parameters, giving
\begin{equation}
\label{L}
\L(M) \equiv - \frac{1}{2} \tilde \chi^2(P_\best)
- \frac{1}{2} \sum_{\mu} \log \lambda_\mu
- \frac{1}{2} \sum_k \log \varsigma^2_k.
\end{equation}
Here $\tilde \chi^2 = \chi^2 + \sum_k P_k^2/\varsigma_k^2$ with
$\chi^2$ defined as in Eq.~(\ref{derivChiSq}),
$P_\best$ represents the maximum likelihood
parameters found in the alternating regression, and $\varsigma^2_k$
are the {\em a priori} parameter variances. Finally, $\lambda_\mu$ are
the eigenvalues of the Hessian of the posterior log-likelihood around
the maximum likelihood value, which we estimate numerically. We then
maximize $\L(M)$ over all possible models in the S-systems structure
to ``select'' a parsimonious model with a high generalization power.

Note that some eigenvectors have $\lambda_\mu\sim
1/\varsigma^2_p=o(1)$. These directions in parameter space are
{\em sloppy} \cite{Gutenkunst:2007gl}, in the sense that they cannot
be inferred from the data because they do not affect the probability
of the data. Thus we define the number of {\em effective}, or {\em
  relevant} parameters in the model relative to the data as the number
of eigenvectors with $\lambda_\mu>1$.  It will be instructive to see
how this number for the ``selected'' model changes as a function of
various properties of the training data.

\section{Tests using a model of yeast glycolysis}

\begin{figure} 
\centering
\includegraphics[scale=0.6]{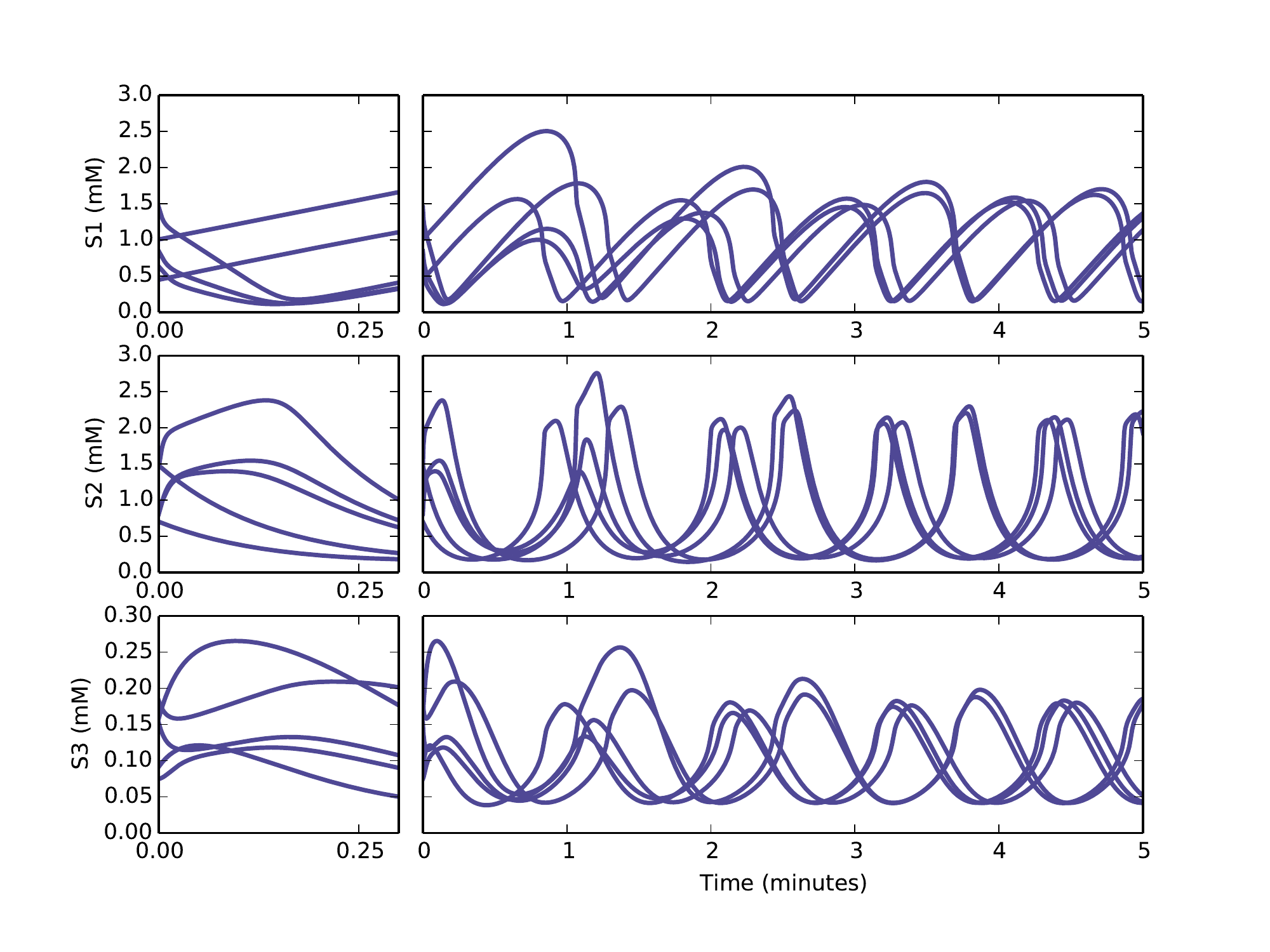}
\caption{\label{yeastTrajectories} Representative trajectories for the
  first three species in the yeast glycolysis model.  Trajectories are
  shown for five initial conditions randomly chosen from the ranges in
  Table~\ref{rangesTable}.  The system's behavior is characterized by
  a stable limit cycle with a period of about one minute, with
  transients that subside on the timescale of seconds to minutes.  At
  left are zoomed portions of the trajectory showing fast
  transients. The period of the oscillation is independent of the
  initial condition, while the phase depends on it.}
\end{figure}

The model of oscillations in yeast glycolysis \cite{RuoChrWol03} has
become a standard test case for biochemical dynamics inference
\cite{Schmidt:2009dt,Daniels:2014vz}. We use it here as well to
analyze the performance of our adaptive S-systems approach.  The
biochemical details are not critical to our current purpose; we
instead take the model as an example of complicated nonlinear dynamics
typical of biological systems. The model consists of ODEs for the
concentrations of 7 biochemical species:
\begin{eqnarray}
\frac{d S_1}{d t} &=& J_0 
    - \frac{k_1 S_1 S_6}{1 + (S_6/K_1)^q}, \nonumber \\
\frac{d S_2}{d t} &=& 2 \frac{k_1 S_1 S_6}{1 + (S_6/K_1)^q}
    - k_2 S_2 (N-S_5) - k_6 S_2 S_5, \nonumber \\
\frac{d S_3}{d t} &=& k_2 S_2 (N-S_5) 
    - k_3 S_3 (A - S_6), \nonumber \\
\frac{d S_4}{d t} &=& k_3 S_3 (A - S_6) - k_4 S_4 S_5 
    - \kappa (S_4 - S_7), \label{yeastEqns} \\
\frac{d S_5}{d t} &=& k_2 S_2 (N-S_5) - k_4 S_4 S_5
    - k_6 S_2 S_5, \nonumber \\
\frac{d S_6}{d t} &=& -2 \frac{k_1 S_1 S_6}{1 + (S_6/K_1)^q} 
    + 2 k_3 S_3 (A - S_6) - k_5 S_6, \nonumber \\
\frac{d S_7}{d t} &=& \psi \kappa (S_4 - S_7) - k S_7, \nonumber
\end{eqnarray}
where the parameters for the model are taken from \cite{RuoChrWol03}
and listed in Table~\ref{yeastParamsTable}. Representative
trajectories from this dynamical system are displayed in
\figref{yeastTrajectories}. Notice that, for the first few minutes,
the system explores a large volume in the phase space, and it settles
down onto a much simpler limit-cycle behavior a few minutes into the
process.

\begin{table}
  \caption{\label{yeastParamsTable}
    Parameters for the yeast glycolysis model 
    defined in Eqs.~(\ref{yeastEqns}), from
    \cite{RuoChrWol03}.}
\begin{center}
\begin{tabular}{ c | l l | | c | l l }
  \hline
  $J_0$   & 2.5    & mM min$^{-1}$          &
  $k$     & 1.8    & min$^{-1}$              \\
  $k_1$   & 100.   & mM$^{-1}$ min$^{-1}$   &
  $\kappa$& 13.    & min$^{-1}$              \\
  $k_2$   & 6.     & mM$^{-1}$ min$^{-1}$   &
  $q$     & 4      &                         \\
  $k_3$   & 16.    & mM$^{-1}$ min$^{-1}$   &
  $K_1$   & 0.52   & mM                      \\
  $k_4$   & 100.   & mM$^{-1}$ min$^{-1}$   &
  $\psi$  & 0.1    &                         \\
  $k_5$   & 1.28   & min$^{-1}$             &
  $N$     & 1.     & mM                      \\
  $k_6$   & 12.    & mM$^{-1}$ min$^{-1}$   &
  $A$     & 4.     & mM                      \\
\end{tabular}
\end{center}
\end{table}

\begin{table}
  \caption{\label{rangesTable}
    Ranges of initial conditions 
    (matching 
    \cite{Schmidt:2011gf})
    and the experimental noise
    (chosen as $1/2$ the standard deviation
    of derivatives sampled from the long-time
    behavior of each species) for training data.
    In test data,
    the high ends of initial condition ranges
    are changed to increase the size of each
    range by $25\%$, forcing the fits to extrapolate, and not just to interpolate.
  }
\begin{center}
\begin{tabular}{  c | c | c }
  {Species}
  &{IC Range} (mM) 
  &{$\sigma_{x'}$} (mM/min) \\
  \hline                        
  $S_1$   & [0.15, 1.60]       
  & 0.2436              \\
  $S_2$   & [0.19, 2.16]       
  & 0.3132              \\
  $S_3$   & [0.04, 0.20]       
  & 0.0252              \\
  $S_4$   & [0.10, 0.35]       
  & 0.0407              \\
  $S_5$   & [0.08, 0.30]       
  & 0.0190              \\
  $S_6$   & [0.14, 2.67]       
  & 0.3739              \\
  $S_7$   & [0.05, 0.10]       
  & 0.0080              \\
\end{tabular}
\end{center}
\end{table}

We evaluate the performance of three types of expansions of ${\mathbf
  X}'$ for the yeast oscillator: a linear representation,
Eq.~\ref{eq:linear}, a fully-connected S-system with all $2J(J+1) =
112$ parameters fit, and an adaptive S-system with a number of fit
parameters that depends on the given data.  To create training data,
we choose $N$ random initial conditions uniformly from ranges shown in
Table~\ref{rangesTable} \cite{Schmidt:2011gf}, integrate each forward
in time using Eqs.~(\ref{yeastEqns}) until time $t_i$ chosen uniformly
from time 0 to time $T$, and use the values of $\vecx$ and $\vecx'$ at
$t_i$ as the input data. To simulate experimental noise, we corrupt
each evaluated rate by a rather large Gaussian noise with the standard
deviation of $0.5\sigma_{{x'}_\mu}$, where $\sigma_{{x'}_\mu}$ is the
standard deviation of the rate $x'_\mu$ sampled over long-time
behavior of each species, also shown in Table~\ref{rangesTable}.
Finally, to evaluate performance in
predicting derivatives, we create out-of-sample test data using the
same method, with timepoints ranging from $t = 0$ to $5$ minutes. For
testing, we force all inference methods to extrapolate rather than
interpolate by choosing a wider range of initial conditions, as
described in Table~\ref{rangesTable}. In this setup, the difficulty of
the inference problem is adjusted by varying $T$.  Indeed, for $T=5$,
much of the training data is close to the low-dimensional
attractor. Since the test data is very similar, the inference only
needs to learn the structure of the derivatives near the attractor in
this case. In contrast, small $T$ forces the algorithms to approximate
the system over a larger volume of the phase space, which is harder
since nonlinearities start playing a stronger role.

The top left plot in \figref{performancePlot} displays the performance
of each method in the easiest case (with $T=5$ minutes), as measured
by the mean correlation between predicted and actual out-of-sample
derivatives, and as a function of the number of data points $N$ in the
training set.  In this case, much of the training/testing dynamics
falls onto the simple, low-dimensional attractor
(cf.~\figref{yeastTrajectories}), and the resulting oscillations are
well-fit by the linear regression, though the adaptive S-system method
performs nearly as well, beating the linear approach around
$N=10$. On the contrary, since the simple dynamics near the attractor
does not require a complex model, the fully connected S-system
overfits and performs significantly worse. Indeed, the linear
regression and the adapted S-system make their predictions with far
fewer parameters than the fully connected model (cf.~top right panel in
\figref{performancePlot}). For comparison, notice also that the
approaches requiring identification of a true microscopic model using
variable and rate time series \cite{Schmidt:2011gf} or building a
phenomenological model using only variable time series
\cite{Daniels:2014vz} both required orders of magnitude more data and
computing time to achieve the same level of out-of-sample correlation
of $0.6$ to $0.8$.

Increasing the difficulty of the inference problem by decreasing $T$
forces approximating the system over a broader, more nonlinear
range. This more severely degrades the predictions of the simple
linear regression than those of the S-systems, as displayed in the
bottom two rows of \figref{performancePlot}.  With $T=0.1$ minutes,
the selected S-system is the best performer, slightly better than the
fully connected S-system, and with significantly fewer nominal and
effective parameters.  With $T = 0$, the number of parameters used in
the selected S-system approaches the full model. Both are equivalent
and perform better than the simple linear regression, as we expect if
they are better able to fit the nonlinearities present in the original
model.

\section{Discussion}

\begin{figure}
\centering
\includegraphics[scale=0.55]{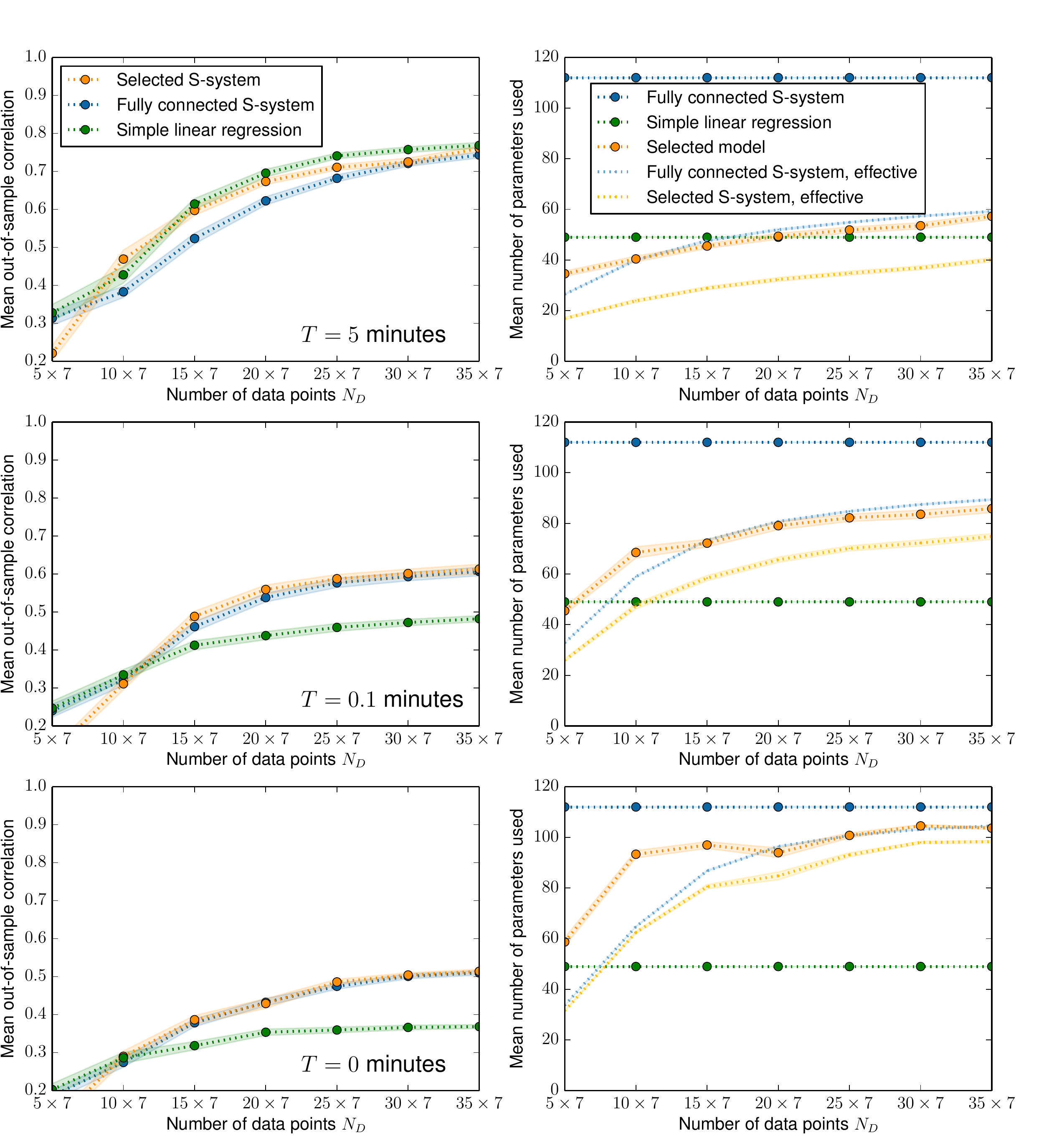}
\caption{\label{performancePlot}
  Inference results for yeast oscillator model.  (Left column) Mean
  correlation of predicted and true derivative values on out-of-sample
  test data versus the amount of data, $N$ (since there are 7 species,
  the total number of data points is $7\times N$), for each of three
  inference methods.  The mean is taken over the 7 species and 100
  different realizations of the training data.  Shaded areas represent
  the standard deviation of the mean over the realizations.  (Right
  column) The nominal and the effective (that is, stiff, or determined
  by the data) number of parameters used in each fit.  The fully
  connected S-system and simple linear regression each have a constant
  nominal number of parameters, while the nominal number of parameters
  in the selected S-system adapts to the data. The effective number of
  parameters is always data dependent. Rows correspond to values of
  $T$, the upper limit on the range of time values used in training
  (test data always use $T = 5$ minutes). As described in the text,
  the complexity of the inference task is higher for lower $T$.  In
  each case, the adaptive S-system model performs at least as well as
  the other approaches. Nonlinearity of S-systems makes them usable
  even for the complex inference task, where the range of variables is
  large and the nonlinear effects are important, so that the linear
  regression model fails ($T=0$ min).  }
\end{figure}

Our results confirm that the alternating regression approach for
inferring phenomenological, approximate dynamics using S-systems
retains the computational efficiency of simple linear regression,
while including nonlinearities that allow it to make better
predictions in a typical test case from systems biology.  In addition,
we generalize the approach of \cite{Chou:2006dh} to include
adaptive dynamical model selection \cite{Daniels:2014vz}, which allows
inference of parsimonious S-systems with an optimal complexity.  When
the dynamics are more simple (as in the cases with $T>0$ in
\figref{performancePlot}), this allows the S-system to use far fewer
parameters and avoid overfitting.  When dynamics become more nonlinear
(as in the case with $T=0$ in \figref{performancePlot}), the approach
makes use of more parameters to obtain better predictions than a
simple linear model.

Inferring the true microscopic structure of the dynamics is possible,
but requires $N\sim 10^4$ measurements for the glycolysis model
\cite{Schmidt:2011gf}. Here we demonstrated that, when the structure
is unknown {\em a priori} and cannot be inferred given the limited
data set size, $N\sim 10^1$, a phenomenological expansion of the
dynamics can still be useful for making predictions about the system's
behavior.  We have described one such representation based on
S-systems that is extremely computationally efficient, more predictive
than linear models through its incorporation of realistic
nonlinearities, and parsimonious through its use of adaptive model
selection.

\begin{acknowledgments}
This research
  was supported in part by the James S.\ McDonnell foundation Grant
  No. 220020321 (I.\ N.), a grant from the John Templeton Foundation
  for the study of complexity (B.\ D.), the Los Alamos National
  Laboratory Directed Research and Development Program (I.\ N.\ and
  B.\ D.), and NSF Grant No. 0904863 (B.\ D.).
\end{acknowledgments}

\bibliographystyle{unsrt}
\bibliography{PredictivePhenomenology}

\end{document}